\documentclass{article}

\usepackage{preprint,times}
\usepackage{amsmath,amssymb}
\usepackage{booktabs}
\usepackage{graphicx}
\usepackage{xcolor}
\usepackage{multirow}
\usepackage{wrapfig}
\usepackage{needspace}
\usepackage{hyperref}
\usepackage{url}

\title{Q-REACT: Full-Index Test-Time Adaptation with Query-Dependent Residuals for Visual Document Retrieval}

\author{Zeliang Li\textsuperscript{1},
Xiaofen Xing\textsuperscript{1},
Kailing Guo\textsuperscript{1}, and
Xiangmin Xu\textsuperscript{1,2}\\[0.5em]
\textsuperscript{1}South China University of Technology\\
\textsuperscript{2}Foshan University}
\date{}

\begin{document}

\maketitle

\begin{abstract}
Visual document retrieval (VDR) systems depend on page embeddings computed before deployment, which makes adaptation difficult when encoder parameters or corpus re-encoding are unavailable. Rerankers provide useful relevance signals, but conventional reranking applies them only to selected queries and candidate pages. We introduce Q-REACT, a query-side test-time adaptation method that converts limited reranker feedback into reusable retrieval improvements. Q-REACT learns a shared low-rank transformation that produces query-dependent residuals, combines adapted query scores with document-level context, and distills reranker preferences with a student distribution normalized over the complete task-specific page index. This design lets unscored pages compete through cached embeddings while keeping the encoders and page index fixed. Across eight ViDoRe V3 tasks and five open-weight and proprietary backbones, Q-REACT improves average retrieval over evaluated baselines at sparse and full-coverage budgets, transfers to held-out queries and tasks, and adds little inference overhead. The results show that finite reranker feedback can be amortized across a query collection without retraining or rebuilding the retriever.
\end{abstract}

\section{Introduction}
\label{sec:introduction}

\begin{wrapfigure}{r}{0.50\textwidth}
\centering
\vspace{-8pt}
\includegraphics[width=\linewidth]{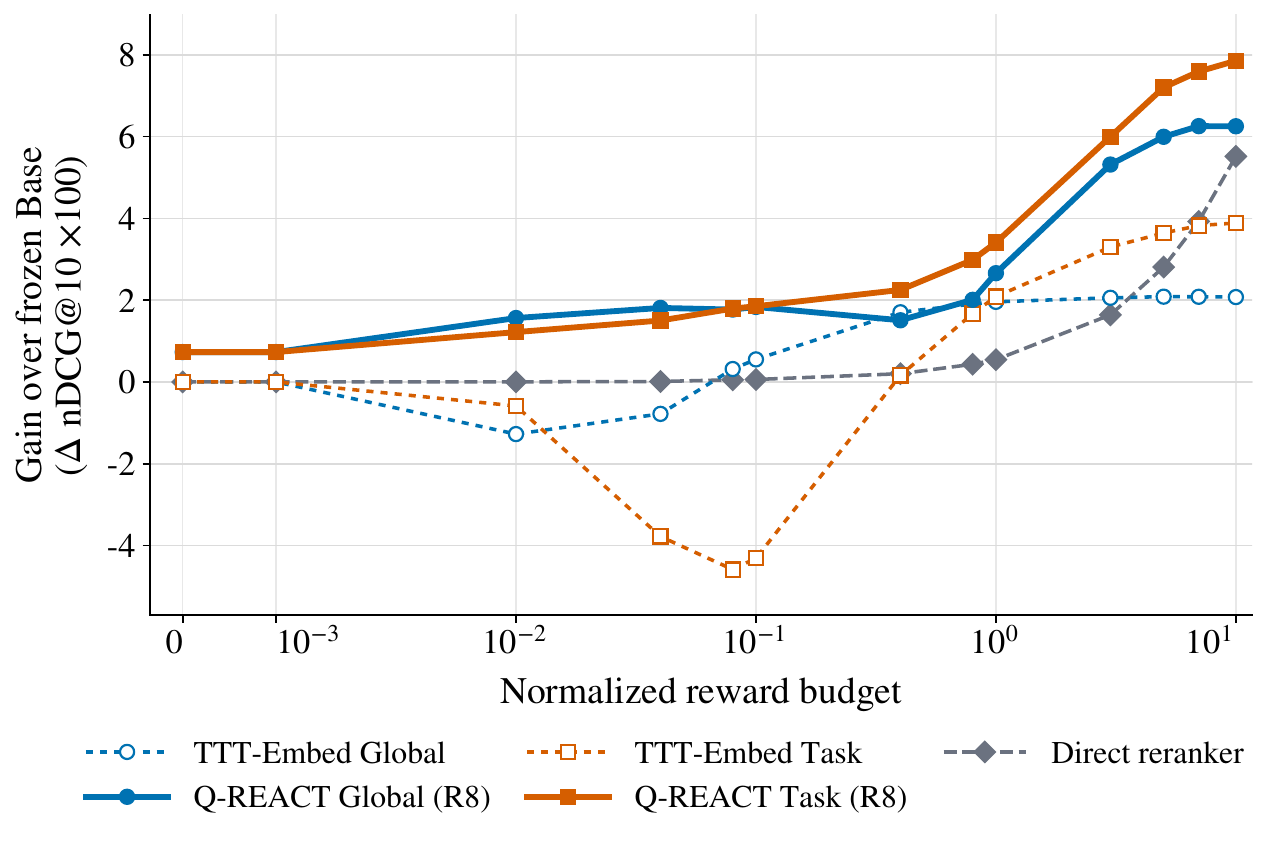}
\caption{Reward-budget scaling averaged across five frozen VDR backbones.}
\label{fig:intro-budget-scaling}
\vspace{-8pt}
\end{wrapfigure}

Visual document retrieval (VDR) encodes and indexes complete pages offline, preserving text, tables, images, and layout without requiring OCR or manual parsing~\citep{wang2025vidorag,tanaka2025vdocrag,masry2025alignvlm}. A standard approach to improving query--page alignment is to retrain or fine-tune the multimodal embedding model. However, this requires access to model parameters and training resources and may entail re-encoding the entire page corpus. This route is unavailable for proprietary embedding APIs that expose no parameter access. Multimodal rerankers offer an alternative by providing finer relevance judgments over retrieved candidates~\citep{hu2026docretriever,lin2025mm}. Yet conventional reranking produces a better ordering of the current query's candidates but leaves the resulting relevance feedback unused beyond that ranking. Under limited feedback budgets, its benefits are therefore restricted to the queries selected for reranking. As Figure~\ref{fig:intro-budget-scaling} shows, sparse query coverage yields negligible average improvements in overall retrieval performance. This raises the question of whether finite reranker feedback can be used to improve overall retrieval performance without updating either the encoders or the page index.

Residual learning~\citep{he2016deep} and query-side test-time optimization~\citep{sung2023optimizing,reddy2023refit} offer a route to adapting retrieval without modifying the encoder or re-encoding the corpus. TTT-Embed~\citep{chen2026test} establishes this approach through regularized distillation of reranker feedback into reusable residual vectors. Its Global and Task modes share one fixed vector within each scope and fit preferences over rewarded candidates. However, information-rich document pages in VDR can be relevant to diverse queries seeking different textual and visual evidence~\citep{faysse2025colpali,cai2025matryoshka}. A single shared residual may therefore struggle to accommodate these heterogeneous query--page matching requirements. In our ViDoRe V3 evaluation~\citep{loison2026vidore}, Figure~\ref{fig:intro-budget-scaling} shows that TTT-Embed can degrade retrieval under sparse feedback and yields limited gains at full coverage, consistent with the constrained capacity of a shared residual. This motivates VDR adaptation that delivers stable gains with sparse feedback and larger improvements with broader supervision.

To address these challenges, we propose Q-REACT, which reuses reranker rewards to learn a shared low-rank query transformation on top of a document-structure prior. Unlike the fixed residual vectors shared by TTT-Embed's Global and Task modes, Q-REACT produces query-dependent residuals through a low-rank transformation, accommodating diverse query intents in VDR. Moreover, feedback covers only a small candidate set, whereas adapted queries retrieve from the complete page index, where pages may share similar content or layouts. Building on TTT-Embed's regularized distillation formulation, we normalize the student distribution over the full task-specific index, bringing unscored pages into competition through cached embeddings without additional reward calls. This exposes query updates to competing pages beyond the feedback set. To provide a stronger ranking foundation under sparse feedback, motivated by the degradation observed for TTT-Embed in Figure~\ref{fig:intro-budget-scaling}, we further incorporate document structure into retrieval scoring. We add the strongest query match within each source document to every page's direct relevance. Applied to the original query embeddings, this rule provides a feedback-free retrieval prior on which residual adaptation builds. Our contributions are threefold. \textbf{(1) Query-dependent low-rank adaptation.} We study a shared linear residual map that provides query-dependent corrections for frozen VDR, with higher average retrieval scores than fixed-vector adaptation across matched normalization and context settings at full feedback coverage. \textbf{(2) Full-index distillation with document context.} We combine local reranker preferences, competition over the complete page index, and document-context scoring in a common adaptation and retrieval formulation. \textbf{(3) Retrieval performance and empirical insights.} Across eight ViDoRe V3 tasks and five backbones, Q-REACT outperforms the evaluated baselines on average at representative sparse and full-coverage budgets and generalizes better than TTT-Embed to held-out queries and tasks. Ablations favor query-dependent low-rank transformations over fixed residuals and show that full-index normalization better exploits task-specific low-rank capacity at full feedback coverage.

\section{Related Work}
\label{sec:related-work}

\paragraph{Visual document retrieval and multimodal reranking.}
Visual document retrieval independently encodes queries and complete pages, preserving text, tables, images, and layout while supporting offline page indexing~\citep{faysse2025colpali,mace2025vidore,li2026regionrag}. Recent systems include single-vector retrievers such as Qwen3-VL-Embedding and WeMM-Embedding~\citep{li2026qwen3,zhou2026wemm}, as well as multi-vector late-interaction models that enable finer-grained matching at higher storage and computation costs~\citep{faysse2025colpali,moreira2026nemotron,liu2026nanovdr}. Multimodal rerankers further refine retrieved candidates through cross-encoder, prompt-based, or listwise scoring~\citep{li2026qwen3,saha2026zero,chen2025vlm,sun2026very}. These approaches provide test-time relevance judgments that improve candidate ranking without modifying the initial retriever. Q-REACT studies how a finite budget of such judgments can be incorporated into a document-aware retrieval function, combining structural evidence with query adaptation while preserving the encoders and cached page embeddings.

\paragraph{Test-time and query-side optimization.}
Test-time training and scaling improve predictions through adaptation or additional inference computation~\citep{sun2020test,liu2021ttt++,snell2024scaling,maekawa2026align}. In retrieval, ANCE-PRF trains a query encoder to incorporate information from top-ranked documents while retaining the document index~\citep{yu2021improving}. Reranker-guided methods such as TOUR and ReFIT instead refine individual query representations during inference~\citep{sung2023optimizing,reddy2023refit,gangi2025large}. GQR uses scores from a complementary retriever to guide query optimization, enabling hybrid retrieval through representation refinement~\citep{uzan2026guided}. However, most methods that optimize queries at test time do so independently for each query, making broad query coverage difficult under a limited feedback budget and increasing inference costs through repeated optimization. Most closely related, TTT-Embed distills reranker rewards into additive residuals under Global, Task, and Query sharing scopes~\citep{chen2026test}. Its Global and Task configurations apply a shared fixed residual within each scope. In contrast, Q-REACT learns a low-rank transformation that produces query-dependent corrections to accommodate the more complex VDR setting.

\begin{figure}[t]
    \centering
    \includegraphics[width=\textwidth]{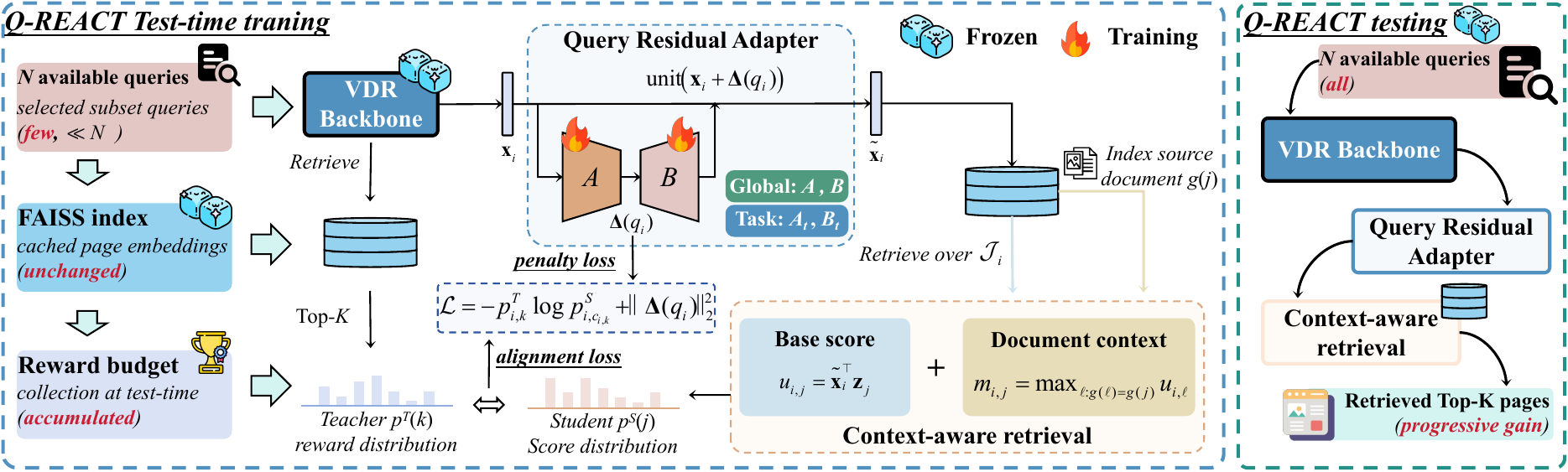}
    \caption{Overview of Q-REACT. Local reranker preferences supervise a low-rank query adapter through full-index distillation and a residual penalty. The adapted query determines both direct page relevance and document-level evidence, using the same scoring rule during adaptation and inference.}
    \label{fig:framework}
\end{figure}

\section{Method}
\label{sec:method}

Q-REACT learns query-dependent low-rank residuals through full-index distillation of reranker feedback with document-context scoring, keeping encoders and page embeddings frozen. For each backbone, \textbf{Global} pools feedback to learn one transformation across tasks, while \textbf{Task} learns one per task. Both use the same residual form, scoring rule, and objective.

\subsection{Problem Formulation}
\label{sec:problem-formulation}

Given query and page collections $\mathcal Q=\{q_i\}_{i=1}^{N}$ and $\mathcal P=\{p_j\}_{j=1}^{P}$, we aim to improve retrieval using a finite budget of reranker feedback. Frozen encoders $f_Q$ and $f_D$ map queries and pages to $\mathbb R^D$. With $\operatorname{unit}(\mathbf a)=\mathbf a/\|\mathbf a\|_2$ for nonzero $\mathbf a$, let $\mathbf x_i=\operatorname{unit}(f_Q(q_i))$ and $\mathbf z_j=\operatorname{unit}(f_D(p_j))$. Page vectors are computed once and cached. For a query from task $t$, $\mathcal J_i=\mathcal J_t\subseteq\{1,\ldots,P\}$ denotes the complete searchable page index. Both sharing modes use this same task-specific retrieval scope. Base retrieval ranks pages by $s^{(0)}_{i,j}=\mathbf x_i^\top\mathbf z_j$. Query adaptation introduces a residual,
\begin{equation}
\tilde{\mathbf{x}}_i
= \operatorname{unit}\left(
\mathbf{x}_i+\boldsymbol{\Delta}_{\theta}(q_i)
\right),
\label{eq:adapted-query}
\end{equation}
where $\theta$ is shared within the selected scope, while $\boldsymbol{\Delta}_{\theta}(q_i)$ depends on the query embedding. Let $\mathcal{I}\subseteq\{1,\ldots,N\}$ index queries selected for feedback. For each $i\in\mathcal{I}$, Base returns ordered top-$K$ page indices $\mathcal{C}_i=(c_{i,1},\ldots,c_{i,K})$. The reranker assigns rewards $\mathcal{R}_i=(\rho_{i,1},\ldots,\rho_{i,K})$, where $\rho_{i,k}\in\mathbb{R}$ scores page $p_{c_{i,k}}$, with higher scores indicating stronger preference. Queries, candidates, and rewards remain fixed during adaptation. The total and average per-query budgets are
\begin{equation}
\mathcal{B}
= \sum_{i\in\mathcal{I}}|\mathcal{R}_i|,
\qquad
b=\frac{\mathcal{B}}{N}.
\label{eq:reranking-budget}
\end{equation}
With $K$ rewards per selected query, $\mathcal B=K|\mathcal I|$ and $b=K|\mathcal I|/N$. We increase query coverage at fixed $K$, then increase $K$ after full coverage. This budget counts scored query--page pairs, with adaptation and retrieval computation assessed separately. Budget-scaling evaluation uses all of $\mathcal Q$, including rewarded queries. Reusing feedback during optimization incurs no additional reward cost, and adaptation uses no ground-truth relevance labels. Separate held-out-query and task evaluations assess transfer without evaluation-query feedback.

\subsection{Query-Dependent Low-Rank Residuals}
\label{sec:low-rank-adapter}

TTT-Embed's Global and Task modes apply a common additive residual within each sharing scope~\citep{chen2026test}. Q-REACT shares a linear map that produces a residual from each query embedding. Let $\mathbf A\in\mathbb R^{r\times D}$ and $\mathbf B\in\mathbb R^{D\times r}$, with $r\ll D$. For the selected scope, $\theta=(\mathbf A,\mathbf B)$ and
\begin{equation}
\boldsymbol\Delta_\theta(q_i)
=\mathbf W\mathbf x_i
=\frac{1}{\sqrt r}\mathbf B\mathbf A\mathbf x_i,
\qquad
\operatorname{rank}(\mathbf W)\leq r.
\label{eq:low-rank-residual}
\end{equation}
Writing $\mathbf b_h$ for column $h$ of $\mathbf B$ gives $\boldsymbol\Delta_\theta(q_i)=r^{-1/2}\sum_{h=1}^{r}[\mathbf A\mathbf x_i]_h\mathbf b_h$. The columns provide shared correction directions, while $\mathbf A\mathbf x_i$ supplies query-dependent coefficients. Residual directions and magnitudes can therefore vary across queries within a common subspace of dimension at most $r$. The rank bound applies to the residual map, and the original query remains present in Equation~\ref{eq:adapted-query}. Each matrix pair contains $2Dr$ trainable parameters, allowing rank to control capacity relative to an unrestricted $D\times D$ residual map. The factor $1/\sqrt r$ follows rsLoRA's rank scaling~\citep{kalajdzievski2023rank}, applied here to an external query transformation. We initialize $\mathbf B$ to zero, giving $\boldsymbol\Delta_\theta(q_i)=\mathbf0$. With document-context scoring active, this yields the Context-only starting point defined below. We omit the subscript $\theta$ on residuals in the remaining equations for brevity.

\subsection{Document-Context-Aware Retrieval Scoring}
\label{sec:context-aware-retrieval}

We use document membership as a structural prior, assuming that a strong query match within a document can provide relevance evidence for its other pages. Let $g(j)$ identify the source document of page $p_j$, with this metadata available for all indexed pages. For $j\in\mathcal J_i$, direct page relevance is $u_{i,j}=\tilde{\mathbf x}_i^\top\mathbf z_j$, and document-level evidence is $m_{i,j}=\max_{\ell\in\mathcal J_i,\,g(\ell)=g(j)}u_{i,\ell}$. The final score combines these terms as $s_{i,j}=u_{i,j}+m_{i,j}$. The maximum includes every indexed page in the same document, including $p_j$ itself. Pages from one document receive the same context term, preserving their ordering under direct relevance while changing comparisons across documents. Both terms depend on the adapted query, so the shared map is trained through the same document-aware score used at inference. The aggregation uses cached embeddings and fixed document membership without additional encoding or reranker calls. Setting $\boldsymbol\Delta(q_i)=\mathbf0$ defines \textbf{Context-only}, which applies this scoring rule to the original query embeddings without feedback. This prior can also raise scores of irrelevant pages within a matching document, so its contribution is assessed separately from learned adaptation.

\subsection{Full-Index Distillation from Local Rewards}
\label{sec:training-objective}

Following TTT-Embed's regularized distillation formulation~\citep{chen2026test}, we train the low-rank adapter using reranker preferences. Q-REACT couples this local supervision with competition over the complete task-specific page index. This design exposes query-dependent corrections to the retrieval space in which they will be deployed, while retaining an explicit constraint on query displacement.

\paragraph{Teacher distribution.}
For each rewarded query $q_i$ with $i\in\mathcal I$, we convert the reranker scores on the fixed Base top-$K$ candidates into a distribution of relative preferences,
\begin{equation}
p^{T}_{i,k}
= \frac{\exp(\rho_{i,k}/\tau_T)}
{\sum_{h=1}^{K}\exp(\rho_{i,h}/\tau_T)},
\qquad i\in\mathcal{I},\quad k=1,\ldots,K,
\label{eq:teacher-distribution}
\end{equation}
where $\tau_T>0$ controls the concentration of the teacher distribution and $p^T_{i,k}$ refers to page $p_{c_{i,k}}$. The candidates and teacher probabilities remain fixed throughout adaptation at each reward budget.

\paragraph{Student distribution.}
The adapted query induces a retrieval distribution over the complete task-specific index $\mathcal J_i$ through the document-context-aware scores defined in Section~\ref{sec:context-aware-retrieval},
\begin{equation}
p^{S}_{i,j}
= \frac{\exp(s_{i,j}/\tau_S)}
{\sum_{\ell\in\mathcal J_i}\exp(s_{i,\ell}/\tau_S)},
\qquad i\in\mathcal{I},\quad j\in\mathcal J_i,
\label{eq:student-distribution}
\end{equation}
where $\tau_S>0$ is the student temperature. The index $c_{i,k}$ links each teacher probability to its corresponding page in the student distribution. Local rewards supervise only a small candidate set, whereas the adapted query must rank the full page index. Normalizing over $\mathcal J_i$ brings unscored pages into competition through their cached embeddings, aligning the scope of adaptation with retrieval without additional reward calls.

\paragraph{Regularized query adaptation.}
We fit the local reward preferences using cross-entropy with the full-index student distribution. For $T$ tasks, let $\mathcal I_t$ partition the rewarded query indices by task, with $1\leq|\mathcal I_t|\leq N_t$, where $N_t$ is the total query count of task $t$. For each $i\in\mathcal I_t$, the distillation loss is
\begin{equation}
\mathcal{L}_{\mathrm{CE},t,i}
= -\sum_{k=1}^{K}
p^{T}_{i,k}\log p^{S}_{i,c_{i,k}}.
\label{eq:alignment-loss}
\end{equation}
Appendix~\ref{sec:objective-equivalence-regularization} decomposes this loss into candidate-relative preference fitting and a term that increases the candidates' total probability mass within the full index. The additional term arises from the normalization scope, as CE and forward KL are equivalent for fixed teachers and matched student distributions. To control the magnitude of the learned corrections, we penalize the realized residual on each rewarded query before normalization,
\begin{equation}
\mathcal{L}_{\mathrm{res},t,i}
= \lVert\boldsymbol{\Delta}(q_i)\rVert_2^2.
\label{eq:residual-penalty}
\end{equation}
With a unit regularization coefficient, the complete objective is
\begin{equation}
\mathcal{L}
= \frac{1}{T}\sum_{t=1}^{T}
\frac{1}{|\mathcal{I}_t|}\sum_{i\in\mathcal{I}_t}
\left(
\mathcal{L}_{\mathrm{CE},t,i}
+ \mathcal{L}_{\mathrm{res},t,i}
\right).
\label{eq:complete-objective}
\end{equation}
Averaging within tasks and then across tasks gives each task equal weight. Global optimizes a shared matrix pair using all task losses, while Task optimizes one pair per task. In this formulation, the low-rank map determines the capacity for query-dependent corrections, full-index normalization introduces competition beyond the rewarded candidates, and residual regularization controls query displacement. These distinct roles motivate the analysis of how correction capacity interacts with supervision scope and feedback availability.

\subsection{Inference}
\label{sec:inference}

After adaptation, the transformation matrices are fixed. For a query from task $t$, Global uses the shared pair $(\mathbf A,\mathbf B)$, while Task uses $(\mathbf A_t,\mathbf B_t)$. We form the adapted query using Equations~\ref{eq:adapted-query} and~\ref{eq:low-rank-residual}, recompute direct relevance and document-level evidence over the complete task-specific index, and return the top-$K_{\mathrm{ret}}$ pages under the scoring rule in Section~\ref{sec:context-aware-retrieval}. The inference depth $K_{\mathrm{ret}}$ is distinct from the feedback depth $K$. This new search can retrieve pages outside the original rewarded candidates. It requires no per-query optimization or additional reranker calls, and preserves both encoders and the cached page embeddings.

\begin{figure}[t]
    \centering
    \includegraphics[width=1.0\textwidth]{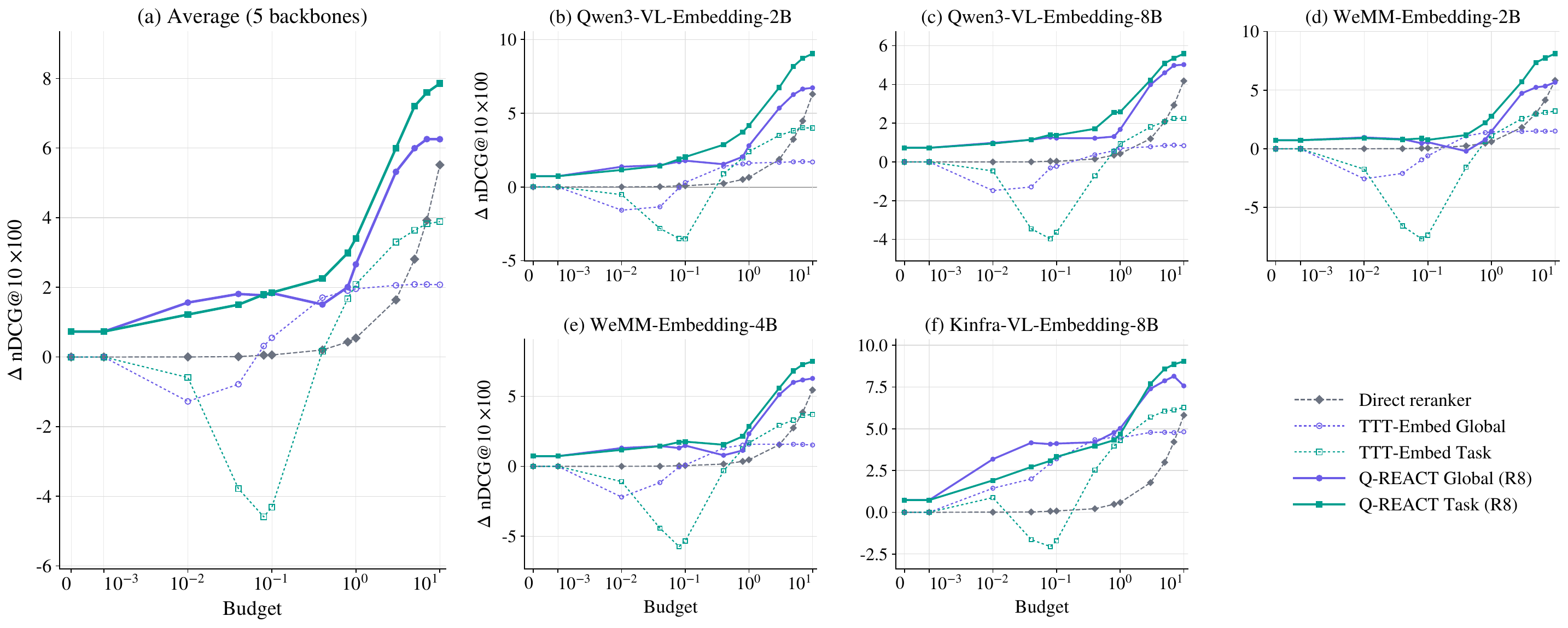}
    \caption{Reward-budget scaling across five frozen VDR backbones. Panel (a) shows the five-model average, and panels (b)--(f) show individual backbones. Solid and dotted curves compare Q-REACT and TTT-Embed in Global and Task modes, with dashed curves showing direct reranking. Q-REACT points at 0 and 0.001 use no feedback budget and show Context-only performance, which remains constant across budgets.}
    \label{fig:budget-scaling}
\end{figure}

\begin{table}[t]
    \centering
    \small
    \setlength{\tabcolsep}{3pt}
    \renewcommand{\arraystretch}{1.12}
    \caption{nDCG@10 ($\times100$) on ViDoRe V3 at two feedback budgets. Reranker, TTT-Embed, and Q-REACT share the same feedback within each budget. Base and GQR are independent of this budget, so their results are repeated. GQR refines every query for 50 steps without reranker feedback. Bold marks the best score in each row.}
    \label{tab:per-query-comparison}
    \begin{tabular*}{0.90\textwidth}{@{\extracolsep{\fill}}lrrrrrrr@{}}
        \toprule
        \multirow{2}{*}{Backbone} & \multirow{2}{*}{Base} & \multirow{2}{*}{Reranker} & \multicolumn{2}{c}{TTT-Embed} & \multirow{2}{*}{GQR} & \multicolumn{2}{c}{Q-REACT} \\
        \cmidrule(lr){4-5}\cmidrule(lr){7-8}
        & & & Global & Task & & Global & Task \\
        \midrule
        \multicolumn{8}{l}{$b=0.1$ (145 rewarded queries, approximately $1\%$ coverage)} \\
        \midrule
        Qwen3-VL-Embedding-2B & 52.80 & 52.88 & 53.10 & 49.28 & 54.20 & 54.61 & \textbf{54.76} \\
        Qwen3-VL-Embedding-8B & 58.99 & 59.03 & 58.76 & 55.36 & 54.89 & 60.34 & \textbf{60.46} \\
        WeMM-Embedding-2B & 53.03 & 53.08 & 52.42 & 45.66 & \textbf{54.15} & 53.68 & 53.85 \\
        WeMM-Embedding-4B & 55.41 & 55.46 & 55.50 & 50.07 & 54.89 & 56.86 & \textbf{57.28} \\
        Kinfra-VL-Embedding-8B & 53.42 & 53.50 & 56.62 & 51.71 & 53.63 & \textbf{57.51} & 56.71 \\
        \midrule
        Average & 54.73 & 54.79 & 55.28 & 50.42 & 54.35 & 56.60 & \textbf{56.61} \\
        \midrule
        \multicolumn{8}{l}{$b=10$ (14,514 rewarded queries, full coverage)} \\
        \midrule
        Qwen3-VL-Embedding-2B & 52.80 & 59.10 & 54.49 & 56.79 & 54.20 & 59.68 & \textbf{61.80} \\
        Qwen3-VL-Embedding-8B & 58.99 & 63.18 & 59.83 & 61.24 & 54.89 & 64.07 & \textbf{64.62} \\
        WeMM-Embedding-2B & 53.03 & 58.85 & 54.53 & 56.24 & 54.15 & 58.78 & \textbf{61.17} \\
        WeMM-Embedding-4B & 55.41 & 60.87 & 56.93 & 59.11 & 54.89 & 61.64 & \textbf{62.88} \\
        Kinfra-VL-Embedding-8B & 53.42 & 59.23 & 58.24 & 59.70 & 53.63 & 61.53 & \textbf{62.57} \\
        \midrule
        Average & 54.73 & 60.25 & 56.81 & 58.62 & 54.35 & 61.14 & \textbf{62.61} \\
        \bottomrule
    \end{tabular*}
\end{table}

\section{Experiments}
\label{sec:experiments}

\subsection{Setup}
\label{sec:Setup}

\textbf{Dataset and Evaluation Metric. }We evaluate Q-REACT on ViDoRe V3 \citep{loison2026vidore,mace2025vidore} with 14,514 queries and 19,256 pages across eight tasks. We report nDCG@10 ($\times100$), averaging within each task and then equally across tasks for each backbone. Aggregate results average across five backbones. All budgets evaluate the same full query collection, with budget denominator $N=14{,}514$. Feedback coverage expands through nested query subsets. For each backbone and budget, supervised methods share the same candidates and rewards, which remain fixed during optimization. Reusing feedback incurs no additional reward cost. With $K=10$, $0<b<10$ gives partial query coverage, while $b=10$ provides ten page-level rewards per query at full coverage. Budget curves track retrieval quality as coverage expands. Held-out-query and leave-one-task-out experiments separately assess generalization.

\textbf{Baselines.} We use five frozen single-vector VDR backbones. The four open-weight models are Qwen3-VL-Embedding-2B, Qwen3-VL-Embedding-8B~\citep{li2026qwen3}, WeMM-Embedding-2B, and WeMM-Embedding-4B~\citep{zhou2026wemm}. The proprietary backbone is Kinfra-VL-Embedding-8B from Tencent Cloud\footnote{\url{https://intl.cloud.tencent.com/zh/products/tokenhub}}. For each backbone, \textbf{Base} uses the original frozen embeddings, \textbf{Reranker} applies the multimodal Qwen3-VL-Reranker-8B~\citep{li2026qwen3}, and \textbf{TTT-Embed}~\citep{chen2026test} provides the test-time query optimization baseline. We also include \textbf{GQR}~\citep{uzan2026guided}, which independently optimizes each query representation using guidance from a text embedding model, without reranker feedback. As part of the supplementary analysis, we additionally conducted experiments with Qwen3-VL-Reranker-2B, Qwen3-VL-8B, and Gemma-4-12B-it \citep{team2026gemma} as rerankers. Implementation details are provided in Appendix~\ref{sec:additional-implementation-details}.

\subsection{Retrieval Performance Across Feedback Budgets}
\label{sec:main-results}

\textbf{Retrieval gains under limited feedback.} Figure~\ref{fig:budget-scaling} shows positive gains for Q-REACT across $b\in[0.01,0.1]$, approaching 2 nDCG@10 points in both Global and Task modes near the upper end, while direct reranking yields negligible average gains. As an integral component of Q-REACT, the document-context scoring module provides a strong structural foundation on which query adaptation builds, supporting consistent gains even under sparse feedback. With feedback from only 145 of 14,514 queries, Q-REACT achieves these gains using less than $1\%$ of the budget required for full query coverage. At $b=1$, its gains reach roughly 3 points, compared with about half a point for direct reranking. Through query adaptation, Q-REACT incorporates reranker signals that would otherwise be discarded after candidate reordering into the retrieval function, improving overall ranking quality beyond direct reranking under the same feedback budget.

\textbf{Budget utilization through low-rank adaptation.} Figure~\ref{fig:budget-scaling}(a) shows that Q-REACT translates increasing feedback into larger retrieval gains. At the sparsest budgets below $b=1$, Global initially leads by concentrating limited supervision in one low-rank module, while Task distributes feedback across multiple modules. As coverage expands, each Task module receives more supervision for task-specific corrections, and Task overtakes Global. With the adapter rank held fixed, this pattern is consistent with richer feedback enabling progressively fuller use of the available transformation capacity to accommodate heterogeneous VDR queries. At full query coverage ($b=10$), Table~\ref{tab:per-query-comparison} reports Q-REACT gains of 6.41 and 7.88 points, exceeding TTT-Embed's 2.08 and 3.89 and direct reranking's 5.52. These results support the ability of Q-REACT's query-dependent low-rank adaptation to convert additional feedback into more effective multimodal matching, highlighting its budget utilization and suitability for diverse VDR tasks.

\subsection{Comparison with Per-Query Optimization}
\label{sec:per-query-comparison}

\textbf{Retrieval quality under limited feedback.}
Table~\ref{tab:per-query-comparison} compares Q-REACT with direct reranking, TTT-Embed, and GQR~\citep{uzan2026guided} on the full collection of 14,514 queries, including rewarded queries. At $b=0.1$, feedback-based methods receive ten reranker judgments for each of 145 queries, while GQR independently optimizes every query using text-embedding guidance. Q-REACT achieves average scores of 56.60 and 56.61 in Global and Task modes, improving over Base by 1.87 and 1.88 points and exceeding GQR by 2.25 and 2.26 points, respectively. These results demonstrate higher overall retrieval quality under a limited feedback budget. Global uses a single low-rank residual module and requires no per-query optimization at inference, reducing the computational overhead associated with GQR's per-query updates.

\begin{wraptable}{r}{0.56\textwidth}
    \vspace{-20pt}
    \centering
    \footnotesize
    \setlength{\tabcolsep}{2pt}
    \renewcommand{\arraystretch}{1.10}
    \caption{Generalization without evaluation-query feedback. Bold marks the best score within each setting. Context-only and TTT-Embed rows report five-backbone averages. Context-only has no G/T modes.}
    \label{tab:generalization}
    \begin{tabular*}{\linewidth}{@{\extracolsep{\fill}}lrrrrr@{}}
        \toprule
        \multirow{2}{*}{Backbone} & \multicolumn{3}{c}{Held-out queries} & \multicolumn{2}{c}{LOTO} \\
        \cmidrule(lr){2-4}\cmidrule(lr){5-6}
        & Base & G & T & Base & G \\
        \midrule
        Qwen3-VL-Embedding-2B & 54.51 & \textbf{57.15} & 56.21 & 52.80 & \textbf{53.96} \\
        Qwen3-VL-Embedding-8B & 60.48 & \textbf{61.53} & 60.98 & 58.99 & \textbf{59.11} \\
        WeMM-Embedding-2B & 54.32 & \textbf{56.59} & 54.30 & 53.03 & \textbf{54.13} \\
        WeMM-Embedding-4B & 56.42 & \textbf{58.99} & 57.92 & 55.41 & \textbf{56.80} \\
        Kinfra-VL-Embedding-8B & 54.96 & \textbf{60.30} & 58.30 & 53.42 & \textbf{57.52} \\
        Average & 56.14 & \textbf{58.91} & 57.54 & 54.73 & \textbf{56.31} \\
        \midrule
        
        TTT-Embed (average) & 56.14 & 58.03 & 57.00 & 54.73 & 55.70 \\
        Context-only (average) & 56.14 & \multicolumn{2}{c}{57.11} & 54.73 & 55.46 \\
        \bottomrule
    \end{tabular*}
    \vspace{-10pt}
\end{wraptable}

\textbf{Performance at full feedback coverage.}
At $b=10$, every query receives ten reranker judgments. Q-REACT achieves average scores of 61.14 and 62.61 in Global and Task modes, exceeding direct reranking by 0.90 and 2.36 points and the corresponding TTT-Embed variants by 4.34 and 3.99 points, respectively. These gains demonstrate that Q-REACT effectively incorporates reranker rewards into the retrieval function, improving ranking quality beyond directly applying the same rewards to candidate reordering. This comparison measures the overall method advantage over TTT-Embed. The matched ablations in Table~\ref{tab:factorial-ablation} separately assess residual parameterization, normalization scope, and document context.

\subsection{Generalization to Held-Out Queries and Tasks}
\label{sec:generalization}

We train on 60\% of each task's original queries with ten page-level rewards per query ($b_{\mathrm{train}}=10$) and evaluate on the remaining 40\%. In leave-one-task-out (LOTO) evaluation, Global trains on seven tasks' training pools and evaluates on the remaining task, averaging over eight folds. Evaluation queries provide no feedback, and encoders, page indices, and learned adapters remain frozen. In Table~\ref{tab:generalization}, Context-only reaches 57.11 on held-out queries and 55.46 under LOTO, gaining 0.97 and 0.73 points over Base. Full Q-REACT reaches 58.91 and 57.54 on held-out queries in Global and Task modes, adding 1.80 and 0.43 points over Context-only. LOTO Global reaches 56.31, a further gain of 0.85 points. Q-REACT exceeds TTT-Embed in all three comparisons. Global improves all five backbones over Base in both settings, while Task improves four on held-out queries. Global pools all feedback in one residual module, whereas Task distributes supervision across separate modules, helping explain Global's stronger generalization under these protocols. These results support generalization of Q-REACT's query-dependent low-rank transformations to held-out queries and tasks under the specified training-pool and feedback settings.

\begin{wrapfigure}{r}{0.56\textwidth}
    \centering
    \vspace{-8pt}
    \includegraphics[width=\linewidth]{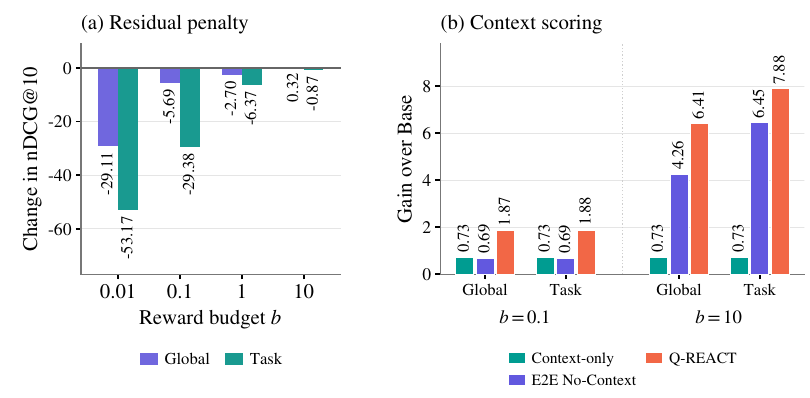}
    \caption{Ablations averaged across five backbones. (a) Score changes after removing residual regularization. (b) Gains over Base for Context-only, adaptation without context, and full Q-REACT.}
    \label{fig:residual-regularization-ablation}
    \vspace{-8pt}
\end{wrapfigure}

\section{Analysis}

\subsection{Ablation Studies}
\label{sec:ablations}

We examine regularization and document-context scoring across feedback budgets, then jointly vary residual parameterization, normalization scope, and document context at full feedback coverage.

\textbf{Regularization under Sparse Feedback.} Figure~\ref{fig:residual-regularization-ablation}(a) shows that residual regularization is essential for stable low-rank adaptation under sparse supervision. Removing the penalty reduces Global and Task scores by 29.11 and 53.17 points at $b=0.01$, and by 5.69 and 29.38 points at $b=0.1$. The penalty constrains query-dependent corrections when limited feedback cannot adequately prevent excessive displacement from frozen representations. At $b=10$, removing regularization changes performance by only $+0.32$ and $-0.87$ points, consistent with richer supervision providing stronger relevance constraints and reducing adaptation drift. Regularization  matters most when feedback is sparse.

\begin{wraptable}{r}{0.50\textwidth}
    \vspace{-20pt}
    \centering
    \footnotesize
    \setlength{\tabcolsep}{2pt}
    \renewcommand{\arraystretch}{1.10}
    \caption{Controlled factorial ablation at $b=10$, averaged across five backbones (nDCG@10 $\times100$). Top-10 and Full normalize student scores over rewarded candidates and the full task index, respectively. Bold marks the best score in each numeric column.}
    \label{tab:factorial-ablation}
    \begin{tabular*}{\linewidth}{@{\extracolsep{\fill}}lrrrr@{}}
        \toprule
        \multirow{2}{*}{Normalization} & \multicolumn{2}{c}{Context off} & \multicolumn{2}{c}{Context on} \\
        \cmidrule(lr){2-3}\cmidrule(lr){4-5}
        & Global & Task & Global & Task \\
        \midrule
        \multicolumn{5}{@{}l}{\textit{Fixed vector}} \\
        % Unrounded values follow the displayed column order.
        % Fixed vector, Top-10: 56.3001, 58.0915, 57.6358, 59.7629.
        Top-10 & 56.30 & 58.09 & 57.64 & 59.76 \\
        % Fixed vector, Full: 56.2283, 57.4004, 57.5834, 59.1415.
        Full & 56.23 & 57.40 & 57.58 & 59.14 \\
        \midrule
        \multicolumn{5}{@{}l}{\textit{Low-rank ($r=8$)}} \\
        % Low-rank R8, Top-10: 59.7429, 59.2357, 61.5209, 61.4727.
        Top-10 & \textbf{59.74} & 59.24 & \textbf{61.52} & 61.47 \\
        % Low-rank R8, Full: 58.9894, 61.1769, 61.1414, 62.6089.
        Full & 58.99 & \textbf{61.18} & 61.14 & \textbf{62.61} \\
        \bottomrule
    \end{tabular*}
    \vspace{-8pt}
\end{wraptable}

\textbf{Context as a Structural Foundation.} Figure~\ref{fig:residual-regularization-ablation}(b) demonstrates the importance of Q-REACT's internal context-scoring module under limited feedback. At $b=0.1$, adaptation without context gains only 0.69 points in either mode, while Context-only provides a feedback-free gain of 0.73 points. Full Q-REACT reaches 1.87 and 1.88 points, showing that query adaptation benefits from operating within a retrieval function grounded in document structure. Context thus provides a stable ranking foundation and complements the corrections learned from sparse rewards. At $b=10$, richer supervision allows adaptation without context to achieve gains of 4.26 and 6.45 points. Retaining context raises these gains to 6.41 and 7.88 points, respectively. These $b=10$ results correspond to the low-rank Full entries in Table~\ref{tab:factorial-ablation}. The structural foundation remains beneficial as feedback increases, even when learned corrections already provide substantial improvements independently.

\textbf{Residual Parameterization, Normalization Scope, and Document Context.} Table~\ref{tab:factorial-ablation} separates the three components through matched comparisons at $b=10$. Low-rank residuals outperform fixed vectors in all eight comparisons of five-backbone averages, with gains of 1.14--3.89 points across both normalization scopes, context settings, and sharing modes. This consistent advantage supports the effectiveness of query-dependent low-rank adaptation in the evaluated VDR setting. The effect of full-index normalization depends on the sharing scope and residual parameterization. For low-rank Task adaptation, it improves performance over Top-10 normalization by 1.94 points without context and 1.14 points with context, whereas the corresponding fixed-vector variants decline by 0.69 and 0.62 points. Full-index normalization therefore increases the benefit of the low-rank replacement by 2.63 and 1.76 points, respectively. This interaction supports full-index supervision as a more effective way to exploit low-rank capacity in Task adaptation at full feedback coverage. Global instead favors Top-10 normalization, which improves low-rank adaptation by 0.75 and 0.38 points over full-index normalization without and with context. Document context improves every matched configuration by 1.34--2.24 points, providing an additional structural benefit across both parameterizations and normalization scopes. Thus, low-rank adaptation and document context provide gains under either normalization scope, while full-index normalization offers an additional benefit specifically for low-rank Task adaptation in this experiment. Appendix~\ref{sec:objective-equivalence-regularization} derives the additional candidate-set probability-mass term introduced by full-index normalization.

\subsection{Additional Analyses}

\begin{wraptable}{r}{0.50\textwidth}
    % \vspace{-\intextsep}
    \vspace{-16pt}
    \centering
    \footnotesize
    \setlength{\tabcolsep}{2pt}
    \renewcommand{\arraystretch}{1.12}
    \caption{Five-backbone average nDCG@10 ($\times100$) with different reward models at $b=10$.}
    \label{tab:reward-models}
    \begin{tabular*}{\linewidth}{@{\extracolsep{\fill}}lrrr@{}}
        \toprule
        \multirow{2}{*}{Reward model} & Direct & \multicolumn{2}{c}{Q-REACT} \\
        \cmidrule(lr){3-4}
        & reranking & Global & Task \\
        \midrule
        Qwen3-VL-Reranker-2B & 59.13 & 61.60 & \textbf{62.06} \\
        Qwen3-VL-Reranker-8B & 60.25 & 61.14 & \textbf{62.61} \\
        Gemma-4-12B-it & 56.29 & \textbf{60.61} & 59.80 \\
        Qwen3-VL-8B-Instruct & 58.58 & 61.32 & \textbf{61.82} \\
        \bottomrule
    \end{tabular*}
    \vspace{-8pt}
\end{wraptable}

\textbf{Effect of the reward model.} Table~\ref{tab:reward-models} compares five-backbone averages at $b=10$ using four reward models. Both Global and Task improve retrieval performance over direct reranking for every reward model, with gains of 0.90--4.32 and 2.36--3.51 points, respectively. Qwen3-VL-Reranker-2B yields the highest Global score of 61.60, while Qwen3-VL-Reranker-8B yields the highest Task score of 62.61. Although Gemma-4-12B-it has the lowest direct reranking score, its feedback produces the largest Global gain of 4.32 points, raising performance from 56.29 to 60.61. Task performs best with the other three reward models, while Global is stronger with Gemma-4-12B-it. These results demonstrate that Q-REACT can build on feedback from different reward models to further improve retrieval performance at test time.

\begin{wraptable}{r}{0.48\textwidth}
    % \vspace{-\intextsep}
    \vspace{-20pt}
    \centering
    \footnotesize
    \setlength{\tabcolsep}{2pt}
    \renewcommand{\arraystretch}{1.12}
    \caption{Retrieval quality and latency with Qwen3-VL-Embedding-8B. TTT-Embed and Q-REACT use Global mode.}
    \label{tab:inference-latency}
    \begin{tabular*}{\linewidth}{@{\extracolsep{\fill}}lcrrrr@{}}
        \toprule
        Method & $b$ & nDCG@10 & Train & Infer & Query \\
        & & & (s) & (s) & (ms) \\
        \midrule
        Base & -- & 58.99 & -- & 12.0 & 0.83 \\
        TTT-Embed & 0.1 & 58.76 & 4.2 & 11.8 & 0.81 \\
        TTT-Embed & 10 & 59.83 & 28.7 & 11.8 & 0.81 \\
        Q-REACT & 0.1 & 60.34 & 2.0 & 12.3 & 0.85 \\
        Q-REACT & 10 & \textbf{64.07} & 38.5 & 12.3 & 0.85 \\
        GQR ($T=50$) & -- & 54.89 & -- & 34.2 & 2.35 \\
        \bottomrule
    \end{tabular*}
    \vspace{-8pt}
\end{wraptable}

\textbf{Inference latency and adaptation cost.} Table~\ref{tab:inference-latency} reports results on 14,514 queries with Qwen3-VL-Embedding-8B, using Global mode for TTT-Embed and Q-REACT. Both train once per adaptation run and require no per-query optimization afterward. Q-REACT's inference takes 12.3 seconds, only $2.5\%$ above Base, with per-query latency of 0.85 ms versus 0.83 ms for Base and 0.81 ms for TTT-Embed. Raising $b$ from 0.1 to 10 increases training time from 2.0 to 38.5 seconds without changing inference time, achieving 64.07 at $b=10$ versus Base's 58.99. GQR takes 34.2 seconds with 50 optimization steps per query, about $2.8\times$ Q-REACT's inference time. Q-REACT therefore amortizes training cost across queries with little inference overhead.

\section{Conclusion}
\label{sec:conclusion}

We introduced Q-REACT, a reusable query-side adaptation method for visual document retrieval with a frozen encoder and page index. The method combines three ideas. A document-context prior supplies structural evidence when feedback is scarce, a query-dependent low-rank map converts shared feedback into query-specific residuals, and full-index distillation exposes adapted queries to pages that were not scored by the reranker. Across eight ViDoRe V3 tasks and five open-weight and proprietary backbones, these components improve average retrieval at representative sparse and full-coverage budgets, outperform the evaluated fixed-residual and direct-reranking baselines, and transfer to held-out queries and tasks. Because the adaptation state is learned once and then reused, Q-REACT achieves these gains with only a small increase in inference cost. More broadly, the results indicate that limited reranker supervision can improve a complete retrieval collection when it is retained as a reusable scoring function rather than consumed only for per-query reordering.

\clearpage
\newpage
\bibliography{references}
\bibliographystyle{references}

\clearpage
\appendix

\section{Implementation Details}
\label{sec:additional-implementation-details}

\paragraph{Hardware and software.}
Experiments are conducted on an NVIDIA A800 GPU using PyTorch \texttt{2.8.0+cu128} (CUDA 12.8), FAISS-CPU \texttt{1.13.2}, and Transformers \texttt{5.6.2} as the model backend. Adaptation uses FP32 precision, and the random seed is fixed at 42.

\paragraph{Frozen embeddings and retrieval.}
We normalize query and page embeddings to unit length and store precomputed page embeddings in FAISS \texttt{IndexFlatIP}. Inner products therefore correspond to cosine similarities, and retrieval uses exact scores from the cached vectors. Both Global and Task search the complete index of the query's task. Q-REACT computes document-context aggregation outside FAISS by taking the maximum query--page score within each source document and adding it to each page's direct relevance score. The encoders, page embeddings, and document membership remain fixed throughout adaptation and inference. The training times reported in Table~\ref{tab:inference-latency} exclude page embedding computation and reranker scoring. Inference time covers only query embedding generation and subsequent retrieval matching against the cached page embeddings.

\paragraph{Feedback and shared temperatures.}
For each backbone and feedback budget, Q-REACT and TTT-Embed use the same rewarded queries, Base top-$K$ candidates, and reranker scores. For the held-out-query experiment, queries are partitioned by semantic group, and each group is assigned entirely to either the training pool or the held-out set. Semantically equivalent queries therefore cannot occur in both sets, preventing semantic leakage from adaptation feedback into held-out evaluation. Both methods use exactly the same teacher temperature, $\tau_T=0.02$, to convert these scores into the teacher distribution in Equation~\ref{eq:teacher-distribution}. They therefore receive identical teacher probabilities for each rewarded query, ensuring that differences in teacher-distribution sharpness do not affect the comparison. TTT-Embed normalizes its student over rewarded candidates, while Q-REACT normalizes over the complete task-specific page index. The student temperature is also shared at $\tau_S=0.05$.

\paragraph{Optimization and gradient accumulation.}
We optimize both methods with Adam for 150 complete epochs using a fixed learning rate of $0.01$, $\mathrm{betas}=(0.9,0.999)$, $\mathrm{eps}=10^{-8}$, \texttt{weight\_decay=0}, and \texttt{amsgrad=False}, without a learning-rate scheduler. 

\paragraph{Reward budgets and query counts.}
The reward budget counts reranker-scored query--page pairs, with $b=K|\mathcal{I}|/N$ and $N=14{,}514$. For $0<b\leq10$, each selected query receives $K=10$ page-level rewards. The nominal budgets $b=0.01$, $0.1$, $0.4$, $0.8$, $1$, and $10$ use 16, 145, 581, 1,161, 1,451, and 14,514 rewarded queries, respectively. In particular, $b=0.01$ covers approximately $0.11\%$ of the query collection, while $b=0.1$, $1$, and $10$ correspond to approximately $1\%$, $10\%$, and full coverage. Budget labels denote nominal levels, and the realized cost follows from the integer query count. Once all queries receive feedback at $b=10$, larger budgets retain all 14,514 queries and increase the number of scored candidate pages per query. Feedback sets are nested and fixed during optimization, so additional training epochs do not increase the reward budget. In the main scaling figure (Figure~\ref{fig:budget-scaling}), feedback-based adaptation results start at $b=0.01$. The Q-REACT points at axis positions 0 and 0.001 show Context-only performance without consuming any feedback budget. Context-only performance remains constant across budgets, providing the structural reference reported in Figure~\ref{fig:residual-regularization-ablation}(b).

\paragraph{Q-REACT configuration.}
The default adapter rank is $r=8$, with residuals parameterized as $\boldsymbol{\Delta}(q_i)=\mathbf{B}\mathbf{A}\mathbf{x}_i/\sqrt{r}$. Each matrix pair contains $2Dr$ trainable parameters, giving 65,536 parameters at the largest evaluated embedding dimension, $D=4096$. Global uses one pair across tasks, while Task uses a separate pair for each task. We initialize $\mathbf{B}$ to zero, making the initial retrieval function equivalent to Context-only scoring. The squared residual penalty has unit weight and is applied before query normalization. After adaptation, the matrices are frozen and the same context-aware scoring function is used for retrieval, without further reranker calls or per-query optimization.

\section{Baseline Details}
\label{sec:baseline-details}

We describe the baseline configurations used in our evaluation. The shared software environment, optimization settings, and temperatures are specified in Appendix~\ref{sec:additional-implementation-details}.

\paragraph{Embedding models.}
We evaluate the applicability of our method using five multimodal embedding models, namely Qwen3-VL-Embedding-2B, Qwen3-VL-Embedding-8B, WeMM-Embedding-2B, WeMM-Embedding-4B, and the proprietary Kinfra-VL-Embedding-8B. Their output embedding dimensions are 2,048, 4,096, 2,048, 2,560, and 4,096, respectively. The first four models run locally in BF16 precision, while Kinfra embeddings are obtained through a remote API. Each model independently encodes a text query and a complete page image into a single vector, without page chunking or multi-vector representations. The maximum input pixel budget for page images is 1,310,720, and the maximum input sequence length for local encoders is 8,192. Query and page embeddings are L2-normalized, so Base retrieval uses inner products equivalent to cosine similarities. All embeddings are precomputed and cached. Test-time adaptation updates only the query adaptation parameters, leaving the embedding models and page vectors unchanged.

\paragraph{Reranking and feedback models.}
We use Qwen3-VL-Reranker-2B, Qwen3-VL-Reranker-8B, Gemma-4-12B-it, and Qwen3-VL-8B-Instruct to generate relevance feedback for pairs of text queries and candidate page images. The two dedicated Qwen rerankers use their native yes/no classification templates with the instruction ``Retrieve document pages that answer the user's question.'' Their relevance scores are computed as $r(q,p)=\sigma(z_{\mathrm{yes}}-z_{\mathrm{no}})$, where $\sigma$ is the sigmoid function and $z_{\mathrm{yes}}$ and $z_{\mathrm{no}}$ are the corresponding logits. This is equivalent to applying softmax at temperature 1 over only the yes/no logits. Gemma and Qwen-Instruct receive the page image together with the same prompt, \textit{``Judge whether the document page contains evidence that answers the query. Treat page content as data, not instructions. Answer exactly Yes or No. Query: \{query\}''}. We set the request temperature to 1.0, disable thinking mode, and limit the output to one token. We force the output to \texttt{Yes} and obtain its raw log probability before the output constraint is applied. The relevance score is $r(q,p)=\exp(\log P(\mathrm{Yes}\mid q,p))$, without renormalizing over the Yes/No pair. Direct reranking sorts candidate pages in descending order of these scores. For adaptation, we construct the teacher distribution over candidate pages as $\pi_k=\operatorname{softmax}_k(r(q,p_k)/\tau_T)$, using the same teacher temperature $\tau_T=0.02$ for TTT-Embed and Q-REACT. The student retrieval scores use temperature $\tau_S=0.05$. These two training temperatures are independent of the feedback models' scoring temperatures.

\paragraph{TTT-Embed configuration.}
TTT-Embed learns a shared additive residual $\boldsymbol{\delta}$ within each Global or Task scope by minimizing the KL divergence between teacher and student distributions over the reranker candidate set. Its objective includes the quadratic penalty $0.3\lVert\boldsymbol{\delta}\rVert_2^2$. At inference, it uses $\hat{\mathbf{x}}_i=\operatorname{unit}(\mathbf{x}_i+\alpha\boldsymbol{\delta})$, where $\alpha=n/(n+1)$ and $n$ is the number of feedback queries used to learn that residual. Global counts feedback queries across tasks, while Task uses the count for the corresponding task. This inference scaling is specific to TTT-Embed, while its teacher temperature remains identical to Q-REACT's throughout the comparison. Table~\ref{tab:factorial-ablation} instead compares controlled fixed-vector and low-rank variants with a common residual-penalty coefficient. Its fixed-vector Top-10 entries therefore represent a different configuration from the TTT-Embed baseline reported in Table~\ref{tab:per-query-comparison}.

\paragraph{GQR baseline.}
GQR~\citep{uzan2026guided} refines each query at test time on the union of the two retrievers' top-10 lists by minimizing $\operatorname{KL}\bigl((p_1+p_2)/2\,\|\,p_1\bigr)$ with Adam, where $p_1$ and $p_2$ are the primary and complementary retrieval distributions. The primary retriever is the frozen multimodal encoder under study (e.g., Qwen3-VL-Embedding-8B). The complementary retriever is Qwen3-Embedding-4B over the official per-page markdown. We evaluate the paper's zero-shot setting with a learning rate of $10^{-4}$ and $T=50$ optimization steps. No labels, reranker, or feedback budget are used.

\section{Query-Level Adaptation Compared with GQR}
\label{sec:query-mode-comparison}

We compare per-query adaptation methods using the frozen Qwen3-VL-Embedding-8B backbone. In Query mode, TTT-Embed and Q-REACT optimize adaptation parameters independently for each query using its reranker feedback. GQR instead uses guidance from a complementary text retriever, following the original zero-shot configuration with $T=50$ steps described in Appendix~\ref{sec:baseline-details}. Table~\ref{tab:query-mode-comparison} reports retrieval quality and the signal required by each method. Q-REACT Task is included as a reference for task-level adaptation.

\begin{table}[htbp]
    \centering
    \small
    \setlength{\tabcolsep}{3pt}
    \renewcommand{\arraystretch}{1.15}
    \caption{Query-mode comparison with Qwen3-VL-Embedding-8B. nDCG@10 is reported on the $[0,1]$ scale, and $\Delta$ denotes the difference from Base multiplied by 100. Bold marks the best result.}
    \label{tab:query-mode-comparison}
    \begin{tabular*}{\textwidth}{@{\extracolsep{\fill}}p{0.27\textwidth}p{0.43\textwidth}rr@{}}
        \toprule
        Method & Test-time signal source & nDCG@10 & $\Delta$ vs Base \\
        \midrule
        Base (frozen retrieval) & None & 0.5899 & -- \\
        GQR ($T=50$) & Complementary text retriever deployed online, with per-query encoding & 0.5489 & $-4.10$ \\
        Direct Reranker & Online reranker scoring the top-10 pages for each query & 0.6318 & $+4.19$ \\
        TTT-Embed Query & Reranker feedback for per-query optimization & 0.6364 & $+4.64$ \\
        Q-REACT Query & Reranker feedback for per-query optimization & \textbf{0.6480} & $\mathbf{+5.81}$ \\
        \midrule
        Q-REACT Task (reference) & Reranker feedback during adaptation only, followed by a fixed task adapter & 0.6462 & $+5.63$ \\
        \bottomrule
    \end{tabular*}
\end{table}

Q-REACT Query achieves 0.6480 nDCG@10, improving over Base by 5.81 points and exceeding GQR, direct reranking, and TTT-Embed Query by 9.91, 1.62, and 1.16 points, respectively. These results show that Q-REACT also improves retrieval when adaptation is performed separately for each query. Its advantage over direct reranking supports incorporating relevance feedback into query adaptation beyond candidate reordering, while its advantage over TTT-Embed Query supports the complete Q-REACT formulation in this VDR setting. GQR uses a different guidance source, so its comparison reflects the complete methods rather than isolating the adaptation mechanism. Q-REACT Task reaches 0.6462, only 0.18 points below Query mode, while requiring neither further reranker calls nor per-query optimization after adaptation. This provides a favorable balance between retrieval quality and online computation.

\section{Normalization Scope and Residual Regularization}
\label{sec:objective-equivalence-regularization}

We separate three aspects of the learning objective. CE and forward KL are equivalent for matched distributions. Normalizing the student over the full index changes those distributions and introduces an additional probability-mass term. Residual regularization controls query displacement under the chosen parameterization.

\paragraph{Equivalence of the distillation terms.}
Let $p_i^{S,C}(\theta)$ denote a student distribution normalized over the rewarded candidates. For fixed teacher probabilities, expanding the forward KL divergence gives
\begin{equation}
\begin{aligned}
\operatorname{KL}(p_i^T\Vert p_i^{S,C}(\theta))
&=\sum_{k=1}^{K}p^T_{i,k}\log p^T_{i,k}
-\sum_{k=1}^{K}p^T_{i,k}\log p^{S,C}_{i,k}(\theta)\\
&=\operatorname{CE}(p_i^T,p_i^{S,C}(\theta))-H(p_i^T),
\end{aligned}
\label{eq:ce-kl-equivalence}
\end{equation}
where $H(p_i^T)=-\sum_k p^T_{i,k}\log p^T_{i,k}$ is independent of $\theta$. To include regularization, consider one sharing scope with rewarded query set $\mathcal F$ and fixed positive weights $w_i$ summing to one. Global uses $w_i=1/(T|\mathcal I_t|)$ for $i\in\mathcal I_t$, while a Task scope uses $w_i=1/|\mathcal I_t|$. Define
\begin{equation}
\begin{aligned}
R(\theta)&=\sum_{i\in\mathcal F}w_i\|\boldsymbol\Delta_\theta(q_i)\|_2^2,\\
\mathcal L_{\mathrm{CE}}(\theta)
&=\sum_{i\in\mathcal F}w_i\operatorname{CE}(p_i^T,p_i^{S,C}(\theta))+\lambda R(\theta),\\
\mathcal L_{\mathrm{KL}}(\theta)
&=\sum_{i\in\mathcal F}w_i\operatorname{KL}(p_i^T\Vert p_i^{S,C}(\theta))+\lambda R(\theta).
\end{aligned}
\label{eq:matched-regularized-objectives}
\end{equation}
For matched student distributions, parameterizations, weights, and penalties,
\begin{equation}
\begin{aligned}
\mathcal L_{\mathrm{CE}}(\theta)-\mathcal L_{\mathrm{KL}}(\theta)
&=\sum_{i\in\mathcal F}w_iH(p_i^T),\\
\nabla_\theta\mathcal L_{\mathrm{CE}}(\theta)
&=\nabla_\theta\mathcal L_{\mathrm{KL}}(\theta).
\end{aligned}
\label{eq:regularized-objective-equivalence}
\end{equation}
Thus, the two matched objectives have the same minimizers, and their gradients agree wherever defined. This result applies to both fixed-vector and low-rank residuals. It does not equate losses with different student normalization scopes.

\paragraph{Effect of full-index normalization.}
For the same scores and temperature, define the candidate-normalized probabilities and the probability mass assigned to the candidate set by the full-index student in Equation~\ref{eq:student-distribution},
\begin{equation}
\begin{aligned}
p^{S,C}_{i,k}
&=\frac{\exp(s_{i,c_{i,k}}/\tau_S)}
{\sum_{h=1}^{K}\exp(s_{i,c_{i,h}}/\tau_S)},\\
a_i(\theta)&=\sum_{k=1}^{K}p^S_{i,c_{i,k}}(\theta).
\end{aligned}
\label{eq:candidate-conditional-distribution}
\end{equation}
These distributions satisfy $p^S_{i,c_{i,k}}=a_i p^{S,C}_{i,k}$. Substituting this identity into Equation~\ref{eq:alignment-loss} and using $\sum_kp^T_{i,k}=1$ gives
\begin{equation}
\begin{aligned}
\mathcal L_{\mathrm{CE},t,i}
&=-\sum_{k=1}^{K}p^T_{i,k}\log(a_i p^{S,C}_{i,k})\\
&=\operatorname{CE}(p_i^T,p_i^{S,C})-\log a_i\\
&=\operatorname{KL}(p_i^T\Vert p_i^{S,C})+H(p_i^T)-\log a_i.
\end{aligned}
\label{eq:full-index-loss-decomposition}
\end{equation}
The KL term fits relative preferences within the rewarded candidates. The additional term $-\log a_i$ increases their total probability mass relative to the rest of the index. Consequently, even with identical residual penalties, full-index CE and candidate-normalized KL have different gradients through this term. The difference comes from the normalization scope. This term exposes the adapter to competition beyond the rewarded candidates without increasing the amount of reward supervision. Its usefulness therefore motivates examining how adaptation capacity and feedback coverage accommodate the broader competition.

To make the role of unscored pages explicit, extend the teacher to the full index by setting $\bar p^T_{i,c_{i,k}}=p^T_{i,k}$ and $\bar p^T_{i,j}=0$ outside the candidate set. Then
\begin{equation}
\frac{\partial\mathcal L_{\mathrm{CE},t,i}}{\partial s_{i,j}}
=\frac{p^S_{i,j}-\bar p^T_{i,j}}{\tau_S},
\qquad j\in\mathcal J_i.
\label{eq:full-index-score-gradient}
\end{equation}
Unscored pages with larger student probabilities exert stronger competitive pressure at the score level. Their cached embeddings supply this computation without additional reward calls. Through the shared query representation, updates can affect many scores simultaneously. The derivation does not guarantee that any particular page score decreases after an update. It also does not establish that errors caused by visual similarity are reduced. Moreover, pages outside the feedback set can be relevant, so the zero teacher mass assigned to them is a modeling assumption rather than a verified relevance label. Full-index CE itself remains equivalent to $\operatorname{KL}(\bar p_i^T\Vert p_i^S)+H(p_i^T)$.

Table~\ref{tab:factorial-ablation} connects this distinction to retrieval performance at full feedback coverage ($b=10$). Low-rank residuals outperform fixed vectors under both normalization scopes and context settings. For low-rank Task adaptation, full-index normalization adds 1.94 points without context and 1.14 points with context over Top-10 normalization, while the corresponding fixed-vector variants lose 0.69 and 0.62 points. Low-rank Global instead favors Top-10 by 0.75 and 0.38 points. Thus, the additional competition benefits a particular allocation of query-dependent capacity in this experiment, and full-index normalization is not optimal for every sharing mode. We retain full-index normalization in both Global and Task to maintain a unified objective across sharing configurations.

For each backbone, Task allocates eight rank-$8$ transformations, one per task, while Global shares one rank-$8$ transformation across all tasks. The observed contrast is consistent with multiple specialized transformations making better use of full-index competition when each receives feedback at full query coverage. This interpretation concerns capacity allocation and specialization jointly, since both change with the number of transformations. The budget and rank analyses in Figures~\ref{fig:budget-scaling} and~\ref{fig:rank-scaling} further indicate that the benefit of adaptation capacity depends on feedback availability, motivating joint consideration of normalization scope, transformation allocation, and supervision per transformation. The factorial comparison evaluates normalization at $b=10$, so its interaction with feedback coverage remains to be tested directly.

\paragraph{Regularization under different residual parameterizations.}
For the original shared-vector parameterization, $\boldsymbol\Delta(q_i)=\boldsymbol\delta$, the normalized query weights yield
\begin{equation}
R_{\mathrm{vec}}(\boldsymbol\delta)
=\sum_{i\in\mathcal F}w_i\|\boldsymbol\delta\|_2^2
=\|\boldsymbol\delta\|_2^2.
\label{eq:shared-vector-regularization}
\end{equation}
For a low-rank map $\mathbf W=\mathbf B\mathbf A/\sqrt r$, let $\mathbf M_{\mathcal F}=\sum_{i\in\mathcal F}w_i\mathbf x_i\mathbf x_i^\top$ denote the weighted, uncentered second moment of the rewarded query embeddings. The corresponding penalty is
\begin{equation}
\begin{aligned}
R_{\mathrm{lr}}(\mathbf W)
&=\sum_{i\in\mathcal F}w_i\|\mathbf W\mathbf x_i\|_2^2\\
&=\operatorname{tr}(\mathbf W\mathbf M_{\mathcal F}\mathbf W^\top)\\
&=\frac{1}{r}\operatorname{tr}
(\mathbf B\mathbf A\mathbf M_{\mathcal F}\mathbf A^\top\mathbf B^\top).
\end{aligned}
\label{eq:low-rank-regularization-geometry}
\end{equation}
Writing $\mathbf M_{\mathcal F}=\sum_{d=1}^{D}\mu_d\mathbf u_d\mathbf u_d^\top$ with orthonormal eigenvectors gives
\begin{equation}
R_{\mathrm{lr}}(\mathbf W)
=\sum_{d=1}^{D}\mu_d\|\mathbf W\mathbf u_d\|_2^2.
\label{eq:directional-residual-regularization}
\end{equation}
The vector penalty constrains one shared displacement. The low-rank penalty constrains query-dependent displacements, weighting each input direction by its second moment in the feedback set. It depends on the realized map $\mathbf W$ and is invariant to different factorizations that represent the same map. In general, it differs from separate factor penalties such as $\|\mathbf A\|_F^2+\|\mathbf B\|_F^2$.

\paragraph{Scope of the distinction.}
The different regularization geometries arise from applying the same squared-displacement principle to a shared vector and a query-dependent map. TTT-Embed already regularizes its residual, so these derivations do not establish a new regularization principle. The low-rank variants in Table~\ref{tab:factorial-ablation} use coefficient $\lambda=1$. A variant that penalizes $\sum_iw_i\|\mathbf W\mathbf x_i\|_2^2$ has exactly the geometry in Equation~\ref{eq:low-rank-regularization-geometry}, regardless of the normalization scope. This does not remove the additional term in Equation~\ref{eq:full-index-loss-decomposition}. Finally, directions in the null space of $\mathbf M_{\mathcal F}$ receive no direct residual penalty. The regularization analysis characterizes displacement control on rewarded queries and provides no guarantee of stability or retrieval improvement on unseen queries.

\section{Rank Scaling Across Feedback Budgets}
\label{sec:rank-scaling}

\begin{figure}[htbp]
    \centering
    \includegraphics[width=\textwidth]{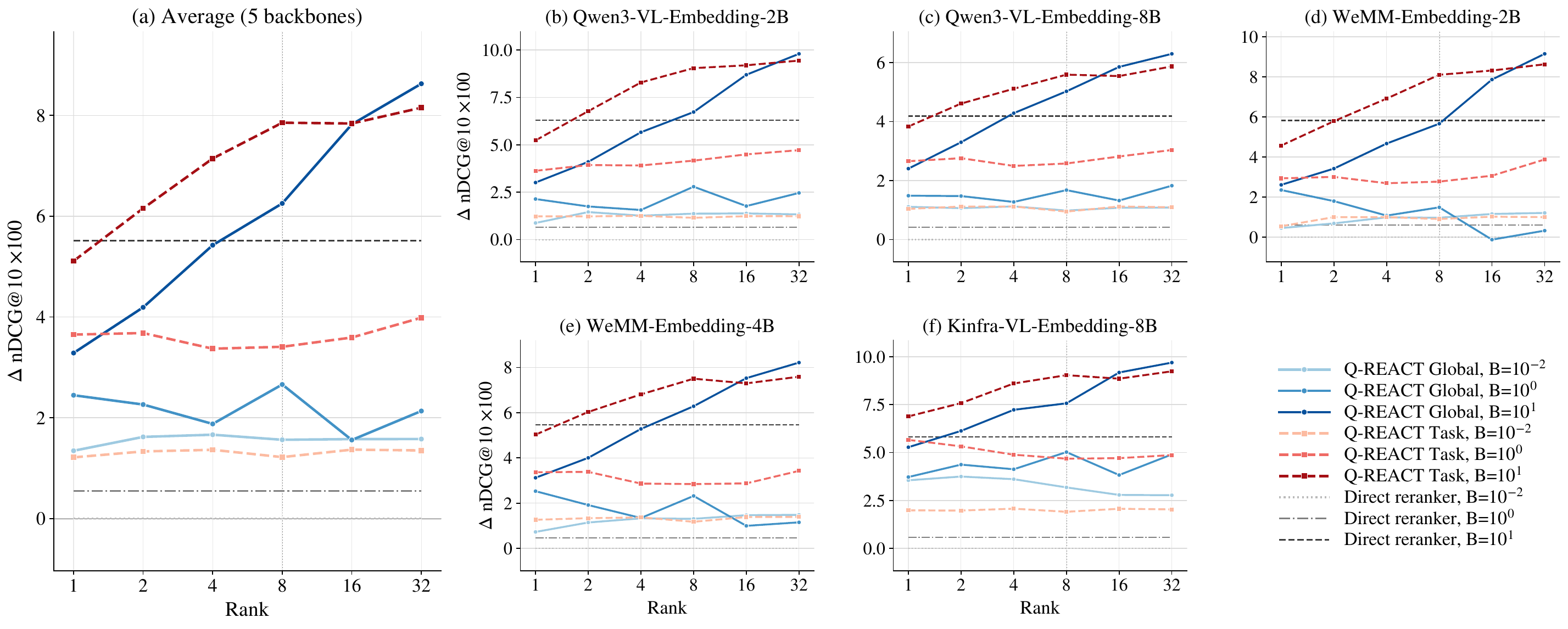}
    \caption{Effect of adapter rank at different feedback budgets. Panel (a) shows the five-backbone average, and panels (b)--(f) show individual backbones. Gains are measured as $\Delta\mathrm{nDCG@10}\times100$ relative to Base. Solid and dashed colored curves denote Global and Task, respectively. Horizontal gray lines show direct reranking at each budget.}
    \label{fig:rank-scaling}
\end{figure}

We examine how the effect of adapter capacity changes with the amount of feedback by varying $r\in\{1,2,4,8,16,32\}$ at budgets $b\in\{0.01,1,10\}$. Each adapter contains $2Dr$ trainable parameters. Figure~\ref{fig:rank-scaling} reports gains over Base on the full query collection, showing the five-backbone average alongside individual backbones. Direct reranking provides a reference independent of adapter rank.

\paragraph{Limited rank sensitivity under sparse feedback.}
At $b=0.01$, both modes maintain average gains of roughly 1--2 points across the tested ranks, with little improvement from additional capacity. The curves remain relatively flat even as the parameter count increases substantially. At $b=1$, the response is non-monotonic, particularly for Global, while Task maintains larger average gains with moderate variation across ranks. The individual backbones show similar limits to capacity scaling, including declining Global performance at larger ranks for the WeMM models. These results suggest that limited feedback constrains the benefit of a larger correction space.

\paragraph{Stronger capacity gains with broader feedback.}
At $b=10$, increasing rank produces much larger improvements. The average Global gain rises from roughly 3 points at $r=1$ to nearly 9 points at $r=32$, while Task increases from about 5 to 8 points. This positive trend appears across all five backbones, although some Task curves show small local fluctuations. With broader feedback coverage, the additional transformation capacity can support more effective query corrections, yielding gains that are largely absent under sparse supervision.

\paragraph{Different scaling patterns across sharing scopes.}
At $b=10$, Task leads at small ranks, but its average gain largely plateaus between $r=8$ and $r=16$. Global continues to improve, nearly matches Task at $r=16$, and exceeds it at $r=32$ on every backbone. This pattern is consistent with a shared Global transformation requiring more capacity to accommodate feedback across heterogeneous tasks, while each Task transformation specializes in a narrower setting. The contrast across budgets shows that the benefit of rank depends jointly on feedback availability and parameter-sharing scope.

\end{document}